%% file: TOP-24-007_temp.tex
\pdfoutput=1
\documentclass[11pt,twoside,a4paper,cmspaper,final,collab]{cms-tdr}
\def\svnVersion{38b2c83}\def\svnDate{2025/08/06}\def\cmsCernNoTag{CERN-EP-2025-061}\def\cmsCernDate{\today}\def\cmsMessage{Published in Reports on Progress in Physics as \href{http://dx.doi.org/10.1088/1361-6633/adf7d3}{\doi{10.1088/1361-6633/adf7d3}.}}
\begin{document}\cmsNoteHeader{TOP-24-007}

\ifthenelse{\boolean{cms@external}}{%
    \providecommand{\cmsTable}[1]{\vspace{1ex}\resizebox{\linewidth}{!}{#1}}
}{%
    \providecommand{\cmsTable}[1]{#1}
}

\newcommand{\ttll}{\ensuremath{\Pell\Pell}\xspace}
\newcommand{\ttlj}{\ensuremath{\Pell\HepParticle{j}{}{}}\xspace}

\newcommand{\mtop}{\ensuremath{m_{\PQt}}\xspace}
\newcommand{\pttop}{\ensuremath{p_{\mathrm{T},\PQt}}\xspace}
\newcommand{\ptatop}{\ensuremath{p_{\mathrm{T},\PAQt}}\xspace}
\newcommand{\mtt}{\ensuremath{m_{\ttbar}}\xspace}
\newcommand{\mll}{\ensuremath{m_{\ttll}}\xspace}
\newcommand{\mbbll}{\ensuremath{\smash{m_{\PQb\PQb\Pell\Pell}}}\xspace}
\newcommand{\chel}{\ensuremath{c_\text{hel}}\xspace}
\newcommand{\chan}{\ensuremath{c_\text{han}}\xspace}

\newcommand{\oneSone}{\ensuremath{{}^1\mathrm{S}_0^{[1]}}\xspace}
\newcommand{\threePone}{\ensuremath{{}^3\mathrm{P}_0^{[1]}}\xspace}

\newcommand{\etat}{{\HepParticle{\PGh}{\PQt}{}}\xspace}
\newcommand{\chit}{{\HepParticle{\PGc}{\PQt}{}}\xspace}
\newcommand{\etatxsec}{\ensuremath{\sigma(\etat)}\xspace}
\newcommand{\etatxsecval}{\ensuremath{8.8\,^{+1.2}_{-1.4}}\xspace}
\newcommand{\etatxsecvalunit}{\ensuremath{\etatxsecval\unit{pb}}\xspace}

\newcommand{\muF}{\ensuremath{\mu_{\mathrm{F}}}\xspace}
\newcommand{\muR}{\ensuremath{\mu_{\mathrm{R}}}\xspace}
\newcommand{\hdamp}{\ensuremath{h_{\text{damp}}}\xspace}
\newcommand{\yt}{\ensuremath{y_{\PQt}}\xspace}

\newcommand{\WW}{\ensuremath{\PW\PW}\xspace}
\newcommand{\WZ}{\ensuremath{\PW\PZ}\xspace}
\newcommand{\ZZ}{\ensuremath{\PZ\PZ}\xspace}
\newcommand{\ttV}{\ensuremath{\ttbar\PV}\xspace}
\newcommand{\tW}{\ensuremath{\PQt\PW}\xspace}
\newcommand{\tX}{\ensuremath{\PQt\mathrm{X}}\xspace}
\newcommand{\Zgst}{\ensuremath{\PZ\text{/}\PGg^\ast}\xspace}
\newcommand{\pp}{\ensuremath{\Pp\Pp}\xspace}
\newcommand{\WbWb}{\ensuremath{\PW\PQb\PW\PQb}\xspace}

\newcommand{\CP}{\ensuremath{CP}\xspace}
\newcommand{\hvq}{\ensuremath{\texttt{hvq}}\xspace}
\newcommand{\bbfourl}{\ensuremath{\texttt{bb4l}}\xspace}
\newcommand{\FXFX}{\ensuremath{\text{FxFx}}\xspace}

\newcommand{\PGnell}{{\HepParticle{\PGn}{\Pell}{}}\xspace}
\newcommand{\PAGnell}{{\HepAntiParticle{\PGn}{\Pell}{}}\xspace}
\newcommand{\Hellp}{{\HepParticle{\widehat{\Pell}}{\PQt}{+}}\xspace}
\newcommand{\Hellm}{{\HepParticle{\widehat{\Pell}}{\PAQt}{-}}\xspace}

\newcommand{\FOpQCDbackground}{{FO pQCD\,+\,back\-ground}\xspace}
\newcommand{\FOpQCDbackgroundetat}{{FO pQCD\,+\,back\-ground\,+\,\etat}\xspace}

\newcommand{\POWHEGhvqPYTHIA}{{\POWHEG v2 \hvq}\,+\,{\PYTHIA}\xspace}
\newcommand{\POWHEGhvqHERWIG}{{\POWHEG v2 \hvq}\,+\,{\HERWIG}\xspace}
\newcommand{\MGvATNLOFXFXPYTHIA}{{\MGvATNLO \FXFX}\,+\,{\PYTHIA}\xspace}
\newcommand{\POWHEGbbfourlPYTHIA}{{\POWHEG vRES \bbfourl}\,+\,{\PYTHIA}\xspace}

\newlength\cmsTabSkip\setlength{\cmsTabSkip}{0.5ex}

\cmsNoteHeader{TOP-24-007}
\title{Observation of a pseudoscalar excess at the top quark pair production threshold}

\date{\today}

\abstract{A search for resonances in top quark pair (\ttbar) production in final states with two charged leptons and multiple jets is presented, based on proton-proton collision data collected by the CMS experiment at the CERN LHC at $\sqrt{s}=13\TeV$, corresponding to 138\fbinv. The analysis explores the invariant mass of the \ttbar system and two angular observables that provide direct access to the correlation of top quark and antiquark spins. A significant excess of events is observed near the kinematic \ttbar threshold compared to the nonresonant production predicted by fixed-order perturbative quantum chromodynamics (pQCD). The observed enhancement is consistent with the production of a color-singlet pseudoscalar (\oneSone) quasi-bound toponium state, as predicted by nonrelativistic quantum chromodynamics. Using a simplified model for \oneSone toponium, the cross section of the excess above the pQCD prediction is measured to be \etatxsecvalunit.}

\hypersetup{%
    pdfauthor={CMS Collaboration},
    pdftitle={Observation of a pseudoscalar excess at the top quark pair production threshold},
    pdfsubject={CMS},
    pdfkeywords={CMS, top, pseudoscalar, scalar, toponium}
}

\maketitle

\section{Introduction}

The discovery of the top quark in 1995 at the Fermilab Tevatron collider was a major milestone in particle physics~\cite{CDF:1995wbb, D0:1995jca}.
Uniquely among quarks, the top quark's lifetime is shorter than the hadronization timescale~\cite{Bigi:1986jk, Bernreuther:2008ju}.
This causes the spin of the top quark to be transferred directly to its decay products, enabling precise measurements of spin properties via angular distributions.
While the individual polarizations of the top quark and antiquark (\PQt and \PAQt) are small when produced via the strong interaction, their spins are correlated in the standard model (SM)~\cite{Bernreuther:2010ny, Mahlon:1995zn}, which was experimentally confirmed at both the Tevatron and the LHC~\cite{D0:2000kns, CDF:2010yag, ATLAS:2012ao, CMS:2013roq, ATLAS:2019zrq, CMS:2019nrx}.

Although \ttbar pairs do not form stable bound states given the short lifetime of the top quark, calculations in nonrelativistic quantum chromodynamics (NRQCD) predict bound state enhancements at the \ttbar threshold~\cite{Fadin:1990wx, Kiyo:2008bv, Sumino:2010bv, Ju:2020otc, Fuks:2021xje, Garzelli:2024uhe, Fuks:2024yjj}.
Since this effect is present only when the \ttbar pairs are in the color singlet configuration, the dominant contribution at the LHC is from the gluon-gluon initial state, leading to the production of the \oneSone ``toponium'' quasi-bound state \etat.
Contributions from other spin states are much smaller at the LHC; for instance, the \threePone state \chit is suppressed by additional powers of the top quark velocity, which is nearly zero at the threshold.
The color octet configuration, on the other hand, is suppressed below the \ttbar threshold because of a repulsive interaction between the top quarks, and has a steeply rising cross section as a function of the \ttbar invariant mass \mtt above the threshold~\cite{Garzelli:2024uhe}.
The presence of such an \etat state would therefore manifest itself as an enhancement in the number of events near the production threshold with distinctive patterns in \ttbar spin correlation observables caused by its pseudoscalar nature.
However, due to the possibility of initial- and final-state radiation, the color configurations of the \ttbar pairs are not necessarily the same as the partons in the initial state, making theoretical predictions of toponium production challenging.

This Letter reports the observation of a threshold enhancement in \ttbar production consistent with pseudoscalar toponium.
The analyzed proton-proton (\pp) collision data at $\sqrt{s}=13\TeV$ were recorded by the CMS experiment at the CERN LHC in 2016--2018, corresponding to an integrated luminosity of 138\fbinv.
The analysis, whose tabulated results are provided in the HEPData record~\cite{hepdata}, is conducted within the context of a search for neutral spin-0 bosons produced through gluon-gluon fusion and decaying to \ttbar.
Here, we focus on the threshold production of a composite \CP-odd pseudoscalar \etat and a \CP-even scalar \chit as signal hypotheses, where \CP refers to the charge-parity symmetry.
These represent the simplest hypotheses that can explain the observation, since they arise naturally within NRQCD.
However, the available experimental data does not exclude alternative explanations like additional pseudoscalar bosons, whose existence is predicted by several theoretical models beyond the SM.
This possibility is explored in Ref.~\cite{CMS:HIG-22-013}, the companion paper to this publication, where the same data is interpreted in terms of limits on additional scalar and pseudoscalar bosons over a large mass range.

The analysis considers final states with two charged leptons (electrons and/or muons) and at least two jets, referred to as the \ttll channel.
A similar analysis was previously performed by the CMS experiment using the data sample collected in 2016 and considering the \ttlj channel (\ie, final states with one charged lepton and at least four jets) in addition to the \ttll channel~\cite{CMS:2019pzc}.
In that analysis, a moderate pseudoscalar-like deviation with a mass at the lowest investigated value of 400\GeV was found.
Compared to that superseded analysis, we consider only the \ttll channel here, but use more than three times the data,
consider resonances with masses below the \ttbar production threshold, and add a second angular observable that provides direct access to
\ttbar spin correlation.

Similar searches have also been conducted by the ATLAS Collaboration using data at $\sqrt{s}=8$~\cite{ATLAS:2017snw} and 13\TeV~\cite{ATLAS:2024vxm}.
The results presented in Ref.~\cite{ATLAS:2024vxm} use the data sample collected in 2015--2018 and combine the \ttll and \ttlj channels, with the latter being predominant.
The analysis in the \ttll channel differs from our approach in that it investigates the invariant mass \mbbll of the $\bbbar\Pellp\Pellm$ system rather than \mtt and it utilizes an angular variable whose sensitivity to \ttbar spin correlation is significantly diluted by kinematic effects.
We have verified that incorporating these differences into our analysis would not result in a significant enhancement at the threshold.
Consequently, the conclusions of Ref.~\cite{ATLAS:2024vxm} are not directly comparable to the ones reported in this paper, nor do they refute or confirm the findings reported herein.

Moreover, our findings are consistent with enhancements at the threshold in previous \ttbar differential cross section measurements reported by ATLAS~\cite{ATLAS:2019zrq, ATLAS:2023gsl} and CMS~\cite{CMS:2019esx, CMS:2019nrx}.
Similarly, the mild tension between the observed and expected measurement of spin correlation in the \ttbar threshold region, which has been reported by both ATLAS~\cite{ATLAS:2023fsd} and CMS~\cite{CMS:2024pts} as part of their studies of quantum entanglement, has been reproduced by this analysis.

\section{Method}

\subsection{The CMS detector and event reconstruction}

The CMS apparatus~\cite{Chatrchyan:2008zzk, CMS:PRF-21-001} is a multipurpose, nearly hermetic detector, designed to trigger on~\cite{Khachatryan:2016bia, CMS:2020cmk} and identify electrons, muons, photons, and charged and neutral hadrons~\cite{CMS:2020uim, CMS:2018rym, CMS:2014pgm, CMS:TRG-19-001}.
A global reconstruction algorithm~\cite{CMS-PRF-14-001} combines the information provided by the all-silicon inner tracker and by the crystal electromagnetic (ECAL) and brass-scintillator hadron calorimeters, operating inside a 3.8\unit{T} superconducting solenoid, with data from gas-ionization muon detectors interleaved with the solenoid return yoke~\cite{CMS:2016lmd, Sirunyan:2019kia, CMS:TRG-19-001}.

{\tolerance=800
Electrons are measured in the pseudorapidity range $\abs{\eta}<2.5$ as energy deposits in the ECAL matched to a track, accounting for their emitted bremsstrahlung energy~\cite{CMS:2020uim, CMS-DP-2020-021}.
Well-identified electron candidates are selected using identification criteria based on boosted decision trees~\cite{CMS:2020uim}.
Muons are measured in the range $\abs{\eta}<2.4$ as tracks in the inner tracker consistent with a track in the muon system.
Well-identified muon candidates are required to pass the ``tight'' working point of the identification criteria described in Ref.~\cite{CMS:2018rym}.
Both electrons and muons are required to be isolated from other activity in the event.
\par}

Hadronic jets are clustered using the anti-\kt algorithm~\cite{Cacciari:2008gp, Cacciari:2011ma} with a distance parameter of 0.4, from particles reconstructed with the particle-flow (PF) algorithm~\cite{CMS-PRF-14-001}.
The charged-hadron subtraction algorithm~\cite{CMS-PRF-14-001} is used to mitigate the effect of additional \pp interactions within the same or nearby bunch crossings (pileup).
Jet energy corrections are applied to match the average measured energy of jets to that of particle-level jets~\cite{CMS:2016lmd}.
Only jets in the range $\abs{\eta}<2.4$ with transverse momentum $\pt>20\GeV$ are considered in the analysis.
Additionally, jets are required to be separated by $\DR=\sqrt{\smash[b]{(\Delta\eta)^2+(\Delta\phi)^2}}>0.4$ from any selected lepton, where $\Delta\eta$ and $\Delta\phi$ are the $\eta$ and azimuthal angle differences between the jet and the lepton.
Jets originating from \PQb quarks are identified (``tagged'') with the \textsc{DeepJet} algorithm~\cite{CMS:2017wtu, Bols:2020bkb, CMS:DP-2023-005}, using the working point with an efficiency of 77\% and a misidentification rate of 15\% for \PQc quark jets and of 2\% for light-quark and gluon jets.
The missing transverse momentum vector \ptvecmiss is computed as the negative vector sum of the \pt of all the reconstructed particles in an event, and its magnitude is denoted as \ptmiss~\cite{Sirunyan:2019kia}.

\subsection{Data and simulated event samples}
\label{sec:samples}

The data used in this analysis were recorded with triggers that require the presence of one or two isolated electrons and/or muons, with an overall trigger efficiency with respect to the offline selection above 96\% in the dielectron channel and 98\% in the dimuon channel.
To compare the collected data to theoretical predictions, Monte Carlo (MC) event samples are generated for all relevant production processes.
Different event generators are used for the calculation of matrix elements (MEs).
In all cases, the next-to-next-to-leading order (NNLO) NNPDF3.1 parton distribution functions (PDFs)~\cite{Ball:2014uwa} are used, and the ME calculation is interfaced with \PYTHIA8.240~\cite{Sjostrand:2014zea} to simulate parton showers, fragmentation, and hadronization, using the CP5 underlying event tune~\cite{Skands:2014pea, CMS:2019csb}.
The top quark mass \mtop is set to 172.5\GeV in all samples involving top quarks, as well as in the computation of theoretical corrections that are applied to them.
The simulated events are processed through the CMS detector simulation based on the \GEANTfour program~\cite{Agostinelli:2002hh}, and pileup interactions generated with \PYTHIA are overlaid to match the observed pileup distribution in data.
Separate MC samples are generated and dedicated corrections to the simulations are applied for different data-taking eras to take into account evolving conditions.

Pseudoscalar \etat production is generated using a simplified model~\cite{Fuks:2021xje, Maltoni:2024tul}, implemented as a generic resonance in the \MGvATNLO 2.6.5 event generator~\cite{Alwall:2014hca} at leading order (LO), with an effective contact interaction describing the gluon-\etat coupling.
Samples of resonant $\etat\to\WbWb$ events are produced, thus including contributions from off-shell top quarks.
The \etat mass and width are set to 343 and 2.8\GeV, respectively, corresponding to the expectation that the toponium mass is twice \mtop minus a binding energy of about 2\GeV~\cite{Fuks:2021xje}, and its decay width is twice the width of the top quark~\cite{Fadin:1990wx}.
The branching ratio of $\PW\to\Pell\PGn$ is taken to be 11.1\%~\cite{Alwall:2014hca}.
The nonrelativistic Hamiltonian reweighting mentioned in Refs.~\cite{Fuks:2021xje, Fuks:2024yjj} is not applied since its impact on the analysis is expected to be negligible~\cite{Sumino:2010bv}.
Scalar \chit production is simulated in the same way with identical mass and width values, but with relevant \CP-odd couplings replaced by \CP-even ones.

The nonresonant \ttbar production process is simulated at next-to-LO (NLO) in perturbative quantum chromodynamics (pQCD) using the \hvq model implemented in \POWHEG v2~\cite{Nason:2004rx, Frixione:2007vw, Alioli:2010xd, Frixione:2007nw}.
The factorization and renormalization scales (\muF and \muR) are set to $\sqrt{\smash[b]{\mtop^2+\pttop^2}}$, where \pttop is the \pt of the top quarks in the underlying Born-level configuration.
Decays of the top quarks are performed using the narrow-width approximation~\cite{Gigg:2008yc}.
The sample is normalized to the predicted \ttbar production cross section of $833.9\,^{+20.5}_{-30.0}\unit{pb}$, as calculated with the \textsc{top++2.0} program at NNLO in pQCD, and including soft-gluon resummation at next-to-next-to-leading-logarithmic (NNLL) accuracy~\cite{Czakon:2011xx}.
To improve the theoretical description of the kinematic distributions of the \ttbar production process, the sample is reweighted to account for higher-order corrections.
The prediction at NNLO in pQCD is calculated using the \textsc{matrix} program~\cite{Grazzini:2017mhc}, and electroweak (EW) corrections at NLO using the \textsc{hathor} program~\cite{Bernreuther:2005is,Kuhn:2005it,Bernreuther:2006vg,Kuhn:2006vh,Aliev:2010zk,Kuhn:2013zoa}.
Both predictions are computed at the level of stable top quarks with a nominal scale choice of $0.5\big(\sqrt{\smash[b]{\mtop^2+\pttop^2}}+\sqrt{\smash[b]{\mtop^2+\ptatop^2}}\big)$, where \ptatop is the \pt of the top antiquark, and using the same NNPDF3.1 PDF set as the \POWHEG\ \ttbar sample.
The \POWHEG\ \ttbar sample is reweighted to the NNLO pQCD and NLO EW predictions using double-differential weights, evaluated at the generator level as a function of \mtt and of the top quark scattering angle with respect to the \ttbar system.
The \ttbar production sample obtained in this way is referred to as fixed-order (FO) pQCD sample in the following.
In addition, three further samples are generated with different generator setups:
(i) the same \POWHEG ME calculation matched to \HERWIG7.2.2~\cite{Bellm:2015jjp, Bahr:2008pv}, using the angular-ordered parton shower model;
(ii) \MGvATNLO at NLO in pQCD with up to two additional jets in the ME and matched to \PYTHIA with the \FXFX merging scheme~\cite{Frederix:2012ps}, using the default scale choices for \FXFX merging; and
(iii) the \bbfourl ME generator in \POWHEG vRES~\cite{Jezo:2015aia, Jezo:2016ujg, Jezo:2023rht}, which simulates the full process $\pp\to\bbbar\Pellp\Pellm\PGn\PAGn$ at NLO in pQCD and thus includes full off-shell contributions beyond the narrow-width approximation, using the default scale choices~\cite{Jezo:2015aia} and matched to \PYTHIA.
The same NNLO pQCD and NLO EW corrections as before are also applied to these samples.
In the case of \bbfourl, the corrections are only applied to the \ttbar part of the full sample. For this purpose, we use ME-based resonance history projectors~\cite{Jezo:2023rht}.

The \etat and \chit samples are added to the FO pQCD \ttbar sample~\cite{Djouadi:2024lyv,Aguilar-Saavedra:2024mnm,Maltoni:2024tul}.
We have tested that adding the \etat and FO pQCD \ttbar samples in this way gives similar results to the most recent NRQCD prediction~\cite{Garzelli:2024uhe}, which uses the NLO QCD potential for Coulomb resummation and includes threshold resummation at next-to-leading-logarithmic (NLL) accuracy.

Reducible background (BG) contributions in the analysis originate from single top quark production (\tX), Drell--Yan production (\Zgst), diboson production (\WW, \WZ, and \ZZ), \ttbar production in association with a vector boson (\ttV), and \ttbar production with an additional misidentified or nonprompt lepton.
The \tX processes via the $t$, \tW, and $s$ channels are generated at NLO using \POWHEG v2, \POWHEG, and \MGvATNLO, respectively~\cite{Re:2010bp, Alioli:2009je,Alwall:2014hca}, and normalized to the NLO cross section predictions for the $t$ and $s$ channels~\cite{Aliev:2010zk, singletopNLO2}, and the approximate NNLO prediction for the \tW channel~\cite{Kidonakis:2021vob}.
To eliminate the overlap between \tW and \ttbar production at NLO, the diagram removal scheme~\cite{Frixione:2008yi} is used.
For the \Zgst process, samples are generated with \POWHEG~\cite{Frixione:2007vw, Alioli:2010xd} with multi-scale-improved NNLO accuracy in QCD~\cite{Monni:2019whf, Monni:2020nks}, matched with \PYTHIA for initial-state radiation (ISR) and \PHOTOS~\cite{Barberio:1993qi, Golonka:2005pn} for final-state radiation (FSR).
The total yield of the \Zgst prediction is corrected using data following a modified version of the procedure described in Ref.~\cite{CMS-TOP-11-002}.
The diboson processes are generated using \PYTHIA and normalized to the respective NNLO (\WW)~\cite{wwNNLO} or NLO (\WZ and \ZZ)~\cite{mcfm} cross sections.
We checked that replacing the \WW contribution with the nonresonant $\PW\PW\bbbar$ production, which leads to the same final state as \ttbar production, does not change the results of this work.
The \ttV events are generated at NLO with \MGvATNLO.
The background contribution due to \ttbar production with an additional misidentified or nonprompt lepton is estimated from the same FO pQCD \ttbar MC sample as described above, considering both the \ttll and \ttlj decay channels of \ttbar.
Background contributions due to {\PW}+jets events with one additional misidentified or nonprompt lepton or due to QCD multijet events with two such leptons are found to be negligible.
Further details on other reducible background predictions are given in Ref.~\cite{CMS:HIG-22-013}.

\subsection{Event selection}

Candidate events are required to have exactly two well-identified leptons (electrons and/or muons) of opposite charge, with $\pt>20\GeV$ and at least one of them with $\pt>25\GeV$.
Events with an additional well-identified electron or muon with $\pt>20\GeV$ are rejected.
The invariant mass \mll of the dilepton pair is required to be larger than 20\GeV to suppress low-mass resonances.
To suppress \Zgst background contributions, events in the $\Pe\Pe$ and $\PGm\PGm$ channels are required to have $\ptmiss>40\GeV$ and $\abs{\mll-91\GeV}>15\GeV$, \ie, to be outside of the \PZ boson mass window~\cite{ParticleDataGroup:2024cfk}.
The presence of at least two jets with $\pt>30\GeV$ and of at least one \PQb-tagged jet with $\pt>20\GeV$ is required, resulting in two or more jets per selected event.

Each event is reconstructed using a kinematic reconstruction algorithm, which assumes that the final state consists of a \ttbar pair that decays into two leptonically decaying \PW bosons.
The algorithm~\cite{CMS-TOP-12-028} proceeds in two steps, first identifying \PQb and \PAQb quark candidates and then reconstructing the momenta of the neutrinos.
In the first step, we consider all \PQb-tagged jets in the event as \PQb and \PAQb quark candidates.
For events with exactly one \PQb-tagged jet, either the \PQb or \PAQb quark candidate, but not both, are taken from the non-\PQb-tagged jets.
The invariant masses of the visible top quark decay products $m_{\Pellp\PQb}$ and $m_{\Pellm\PAQb}$ are calculated for all possible \bbbar candidate pair assignments, and the pair maximizing the likelihood of being the correct pair is chosen for the second step.
The likelihood is constructed using the expected distribution of the invariant mass of the lepton and \PQb-tagged jet system, which has a different shape depending on whether the lepton and \PQb-tagged jet originate from the decay of the same top quark, or not.
The second step uses an analytic method~\cite{Sonnenschein:2006ud} to solve a system of equations in the \PGnell and \PAGnell momenta formed by requiring that:
(i) the invariant masses of the $\Pellp\PGnell\PQb$ and $\Pellm\PAGnell\PAQb$ systems are equal to 172.5\GeV,
(ii) the invariant masses of the $\Pellp\PGnell$ and $\Pellm\PAGnell$ systems are equal to the \PW boson mass, and
(iii) the $\PGnell\PAGnell$ system is the sole source of the measured \ptvecmiss.
In case of multiple real solutions, the one with the lowest reconstructed value of \mtt is used, which has been found to minimize possible biases with respect to the true \mtt value~\cite{CMS:TOP-14-013,Korol:2016wzq}.
This step is repeated 100 times, with random smearing applied to the momenta of the \Pellp, \Pellm, \PQb, and \PAQb candidates, by sampling from the distributions of the relative energy difference and angular distance between reconstructed and truth-level particles.
The effect of the smearing is propagated to the measured \ptvecmiss.
This smearing procedure incorporates the detector resolution into the reconstruction and, in particular, recovers solutions in cases where the nominal configuration would lead to a system of equations without real solution.
The final result of the reconstruction is taken as the weighted averages of the \PQt and \PAQt quark four-momenta, computed over all repetitions that result in a real solution with weight evaluated via the same likelihood based on $m_{\Pellp\PQb}$ and $m_{\Pellm\PAQb}$ used for the \bbbar quark candidate assignment.
Events yielding no real solution in any iteration are rejected.
The resolution of the reconstructed \mtt is around 15\% at the \ttbar threshold and increases up to 25\% for higher \mtt values.
Similarly, the resolution of the reconstructed top quark \pt is around 50\GeV for $\pt<100\GeV$ and increases towards 100\GeV for higher momenta.

Two angular observables \chel and \chan are defined to probe the spin correlation of the \ttbar system as follows.
We define \Hellp and \Hellm as the unit vectors of the momenta of the two leptons in the rest frames of their parent \PQt and \PAQt, respectively~\cite{CMS:HIG-22-013}.
The variable \chel~\cite{Bernreuther:2004jv, Bernreuther:2015yna} is then defined as the scalar product of \Hellp and \Hellm, and \chan~\cite{Aguilar-Saavedra:2022uye} similarly but with a sign flip in the component parallel to the \PQt direction (\ie, the $\widehat{k}$ direction in Ref.~\cite{Bernreuther:2015yna}) of either \Hellp or \Hellm.
The slopes of both distributions are related to the degree of alignment between the \PQt and \PAQt spins.
The production of \etat results in a \ttbar system in the \oneSone state with anticorrelated \PQt and \PAQt spins, resulting in a normalized \chel distribution with a maximal slope of $+1$, steeper than the one of the FO pQCD prediction.
In contrast, \chit production of a \ttbar system in the \threePone state results in a \chel slope of $-1/3$.
Concerning \chan, the normalized distribution predicted by FO pQCD is almost flat, whereas \etat and \chit production result in slopes of $+1/3$ and $-1$, respectively~\cite{CMS:HIG-22-013}.

\subsection{Statistical analysis}

A binned profile maximum likelihood fit~\cite{Junk:1999kv, Read:2002hq, Cowan:2010js, CMS-NOTE-2011-005}, implemented in \textsc{combine}~\cite{CMS:2024onh}, is performed to evaluate the compatibility of the observed data with the predictions and to obtain a best fit estimate of the cross section of the \etat and/or \chit contributions.
A three-dimensional (3D) template built from \mtt, \chel, and \chan is used as the fitted distribution.
Various sources of uncertainty affect the 3D template and are implemented as nuisance parameters in the fit, capable of changing both the rate and the shape of the templates.
Experimental uncertainties related to trigger, electron identification~\cite{CMS:2020uim}, muon identification~\cite{CMS:2018rym}, and \PQb tagging (separately for \PQb/\PQc and \PQu/\PQd/\PQs/\Pg jets)~\cite{CMS:2017wtu} efficiencies, as well as to pileup reweighting, are evaluated by varying the corresponding corrections to the simulation by their uncertainties.
Uncertainties in the jet \pt scale and resolution~\cite{CMS:2016lmd}, as well as an additional uncertainty in the \ptmiss contribution from detector deposits not clustered into jets~\cite{Sirunyan:2019kia}, are evaluated by varying the jet \pt and \ptmiss directly.
Additional nuisance parameters are assigned for an inefficiency caused by the gradual shift in the timing of the ECAL trigger inputs~\cite{CMS:2020cmk}, and for the uncertainty of 1.6\% in the integrated luminosity measurement~\cite{CMS:2021xjt, CMS:2018elu, CMS:2019jhq} that affects the normalization of all simulated processes.
Besides experimental effects, several theoretical uncertainties related to the modeling of the signal and background processes are considered, as described below.

First, we consider theoretical uncertainties in the FO pQCD prediction of \ttbar production.
The NLO EW correction depends on the value of the top quark Yukawa coupling \yt.
The leading \yt dependence is quadratic, originating from the interference of diagrams containing a virtual Higgs boson with LO pQCD diagrams.
A shape uncertainty of $+11/{-}12\%$ in \yt around the nominal value of the SM prediction is evaluated, conservatively using the uncertainty from the measurement reported in Ref.~\cite{CMS:2018uag}.
Following Ref.~\cite{Kuhn:2013zoa}, the difference between the multiplicative and additive application schemes when combining the NLO EW corrections with the NNLO pQCD corrections of about 1--2\% is taken as an additional uncertainty.
We treat the multiplicative application scheme as the nominal EW correction and the additive application scheme as the $+1\sigma$ shift of this uncertainty.
Relative to the nominal template, the $-1\sigma$ shift has the same magnitude and opposite sign as the $+1\sigma$ shift.
All other uncertainties encoding the difference between nominal and alternative modeling choices are implemented similarly.
The uncertainty in \mtop is considered by shifting its value in simulation by $\pm$1\GeV~\cite{Khachatryan:2015hba}.
We have checked that the exact range of the \mtop uncertainty does not influence the conclusions of this work.
The values of \muR and \muF in the ME calculation are varied independently up and down by a factor of 2.
The effects of the \mtop, \muR, and \muF variations on the shape of the 3D template are considered at NLO accuracy, whereas the effects on the overall normalization are taken from the NNLO+NNLL calculation~\cite{Czakon:2011xx, Butterworth:2015oua}.
The scales used to evaluate the strong coupling constant \alpS in the PS simulation of ISR and FSR are varied independently by a factor of 2 in each direction.
Uncertainties in the underlying event tune are estimated by varying the CP5 parameters~\cite{CMS:2019csb}.
Two uncertainties are considered for the color reconnection model, with one based on the ``QCD-inspired'' model~\cite{Christiansen:2015yqa}, and the other by switching on the early resonance decay option in \PYTHIA~\cite{CMS:2022awf}.
The uncertainty in the ME-PS matching scale is evaluated by varying the \POWHEG parameter \hdamp, which controls the radiation of additional high-\pt jets.
The nominal value of \hdamp and the considered variations are $1.58\,^{+0.66}_{-0.59}\,\mtop$~\cite{CMS-PAS-TOP-16-021}.
The uncertainty from the PDF choice is evaluated by reweighting the simulated \ttbar events using 100 replicas of the NNPDF3.1 PDF set, and taking the base variation with the largest eigenvalue as obtained from a principal component analysis in the space of the final 3D templates.
The uncertainty in the value of \alpS in the PDF set is used as a second independent PDF uncertainty.

The choice of ME event generator for the \ttbar FO pQCD prediction is validated through a comparison of the nominal \ttbar FO pQCD + \tW prediction with the alternative prediction obtained with \bbfourl, which includes off-shell contributions to the $\bbbar\Pellp\Pellm\PGn\PAGn$ final state.
The difference between the two predictions is assigned as an additional uncertainty.
Similarly, the comparison of the nominal \ttbar FO pQCD prediction with the alternative generator setup using \HERWIG is used to assign an additional uncertainty for the choice of the parton shower program.

The uncertainties due to \muF, ISR, and FSR scales in the \etat and \chit simulations are evaluated by independently varying them up and down by a factor of 2.
Since the employed model uses effective \etat and \chit production via a contact interaction with no dependence on \alpS, variations of \muR have no effect.
The uncertainty in \mtop is considered by varying its value in simulation by $\pm$1\GeV, and is treated as fully correlated with the \mtop variation in the FO pQCD \ttbar prediction.
Other theoretical uncertainties in the \etat and \chit simulations, such as the PDF variations, are found to be small compared to the already considered effects and are thus neglected.

The \muR, \muF, ISR, and FSR scale uncertainties are also independently considered for the \Zgst and \tX processes.
Normalization uncertainties for the non-\ttbar background processes are assigned based on the precision of relevant cross section measurements~\cite{Aaboud:2016ymp, Sirunyan:2018rlu, Sirunyan:2018lcp, Sirunyan:2017uzs, Aaboud:2019njj, ATLAS:2016oxs}, resulting in larger uncertainties than those of the corresponding theoretical computations.

\section{Results}

\begin{figure*}[!t]
\centering
\includegraphics[width=0.32\textwidth]{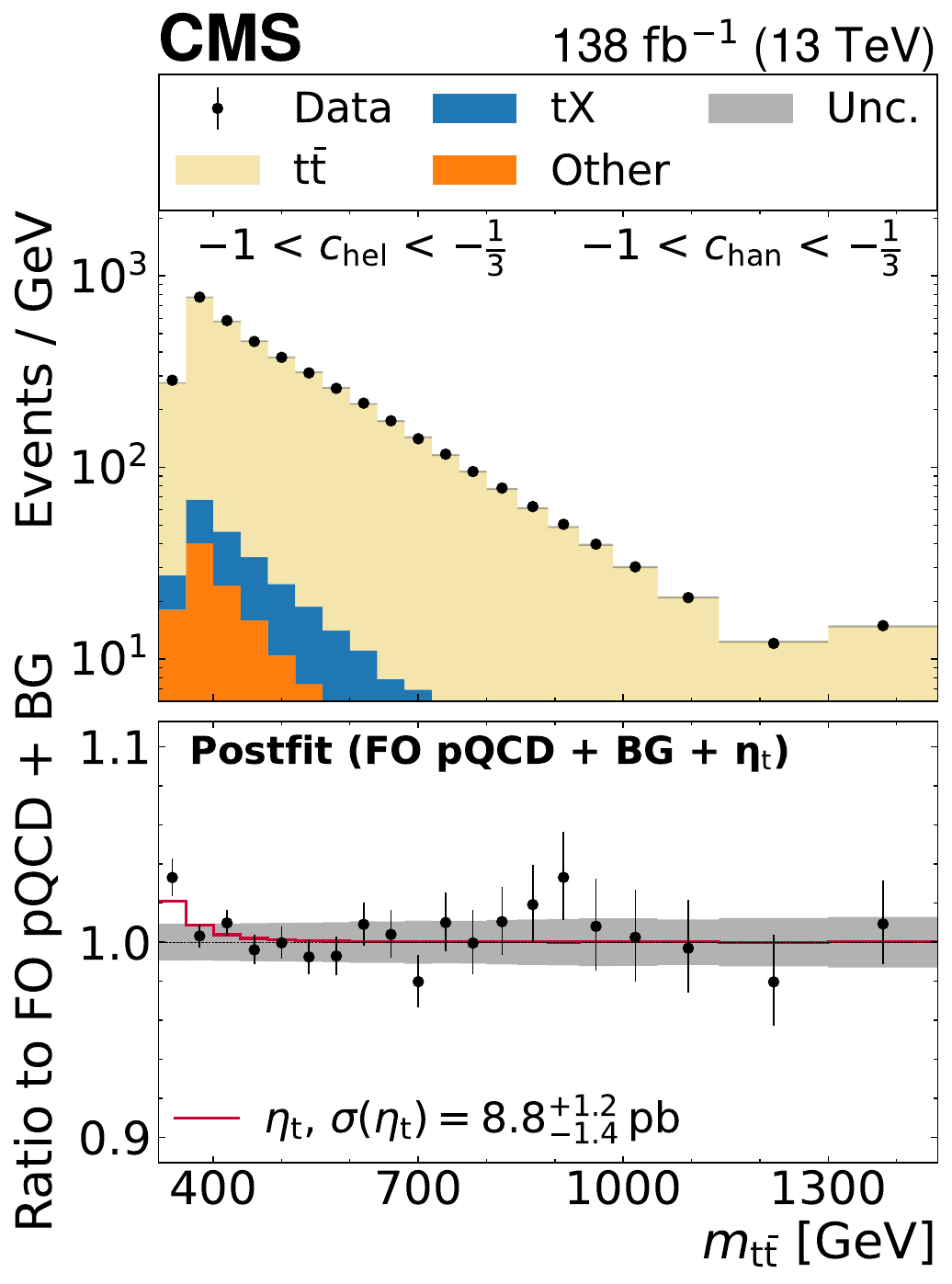}%
\hfill%
\includegraphics[width=0.32\textwidth]{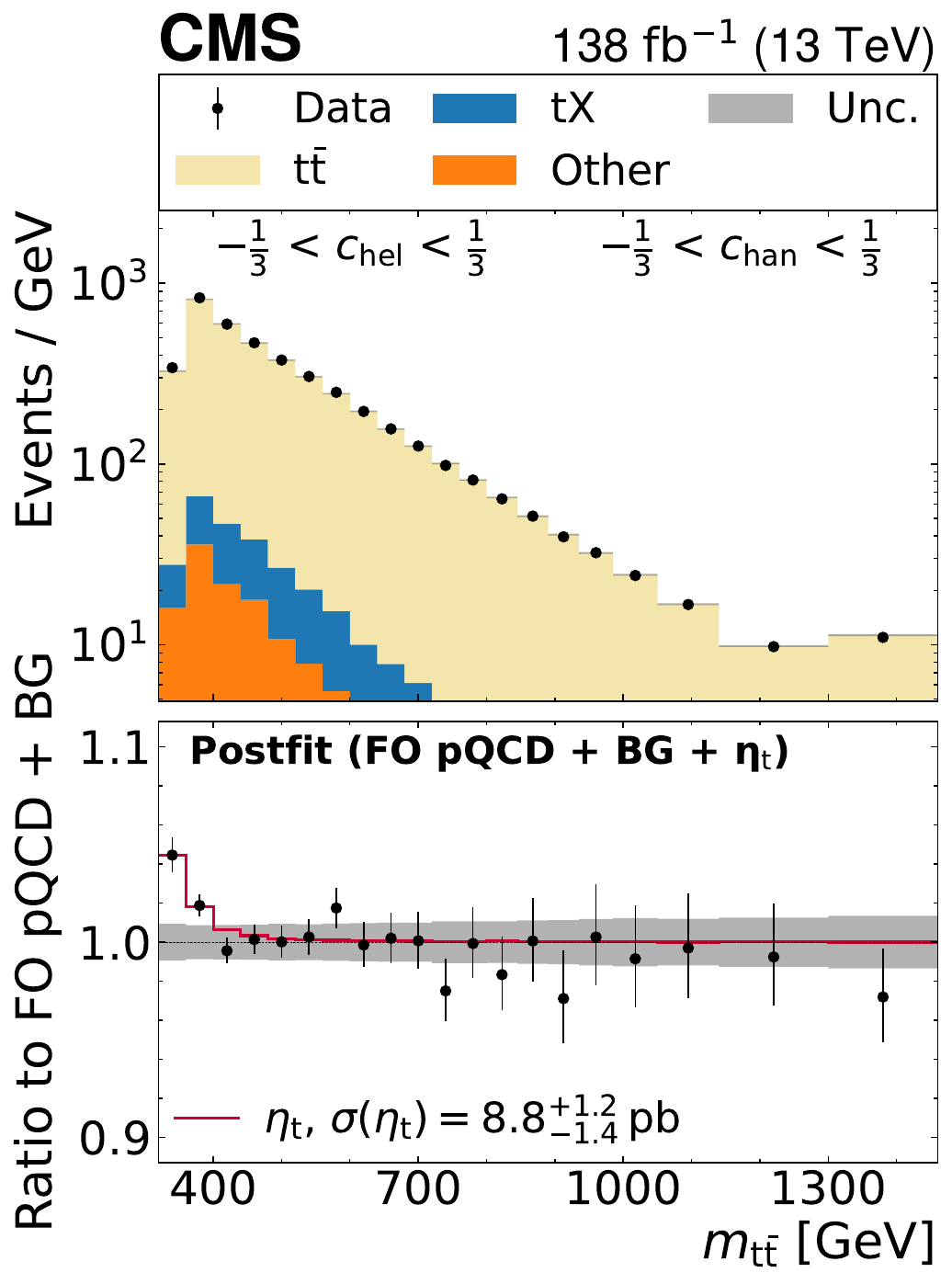}%
\hfill%
\includegraphics[width=0.32\textwidth]{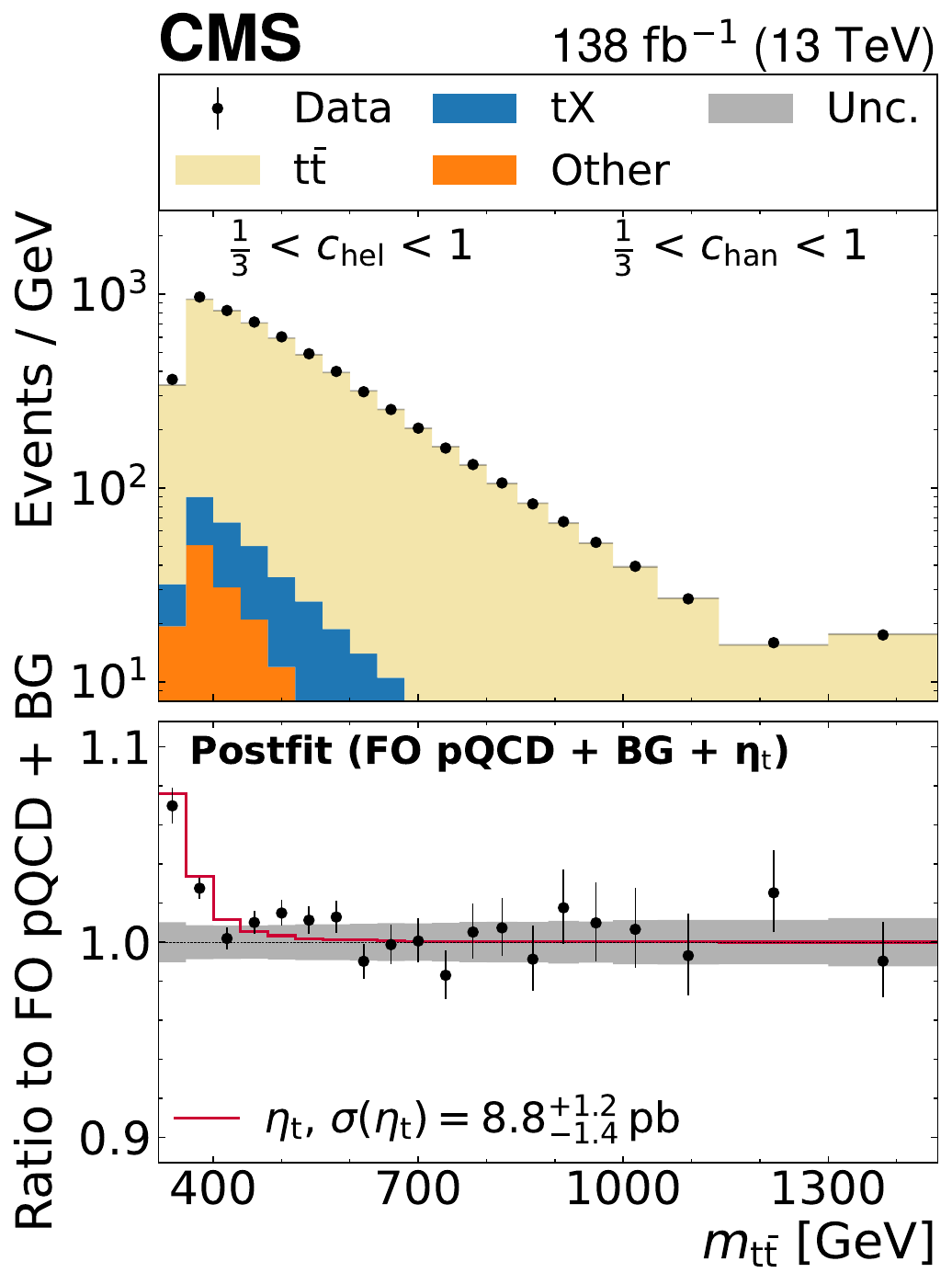}
\caption{%
    Observed (points with statistical error bars) and predicted (stacked colored histograms) \mtt distribution in three out of nine (\chel, \chan) bins.
    In the upper panels, the \ttbar histogram shows the FO pQCD prediction after the fit to the data that includes the \etat signal model (whose contribution is not drawn), and the shown event rates are divided by the bin width.
    The lower panels display the ratio of the data to the \FOpQCDbackground prediction, with \etat signal overlaid at its best fit \etat cross section (red line).
    The gray band indicates the postfit uncertainty.
    The first and last \mtt bins include all events with reconstructed \mtt below 360 and above 1300\GeV, respectively, and the drawn bin width is used for the normalization in these bins.
}
\label{fig:combined_ll}
\end{figure*}

\begin{figure*}[!htp]
\centering
\includegraphics[width=0.32\textwidth]{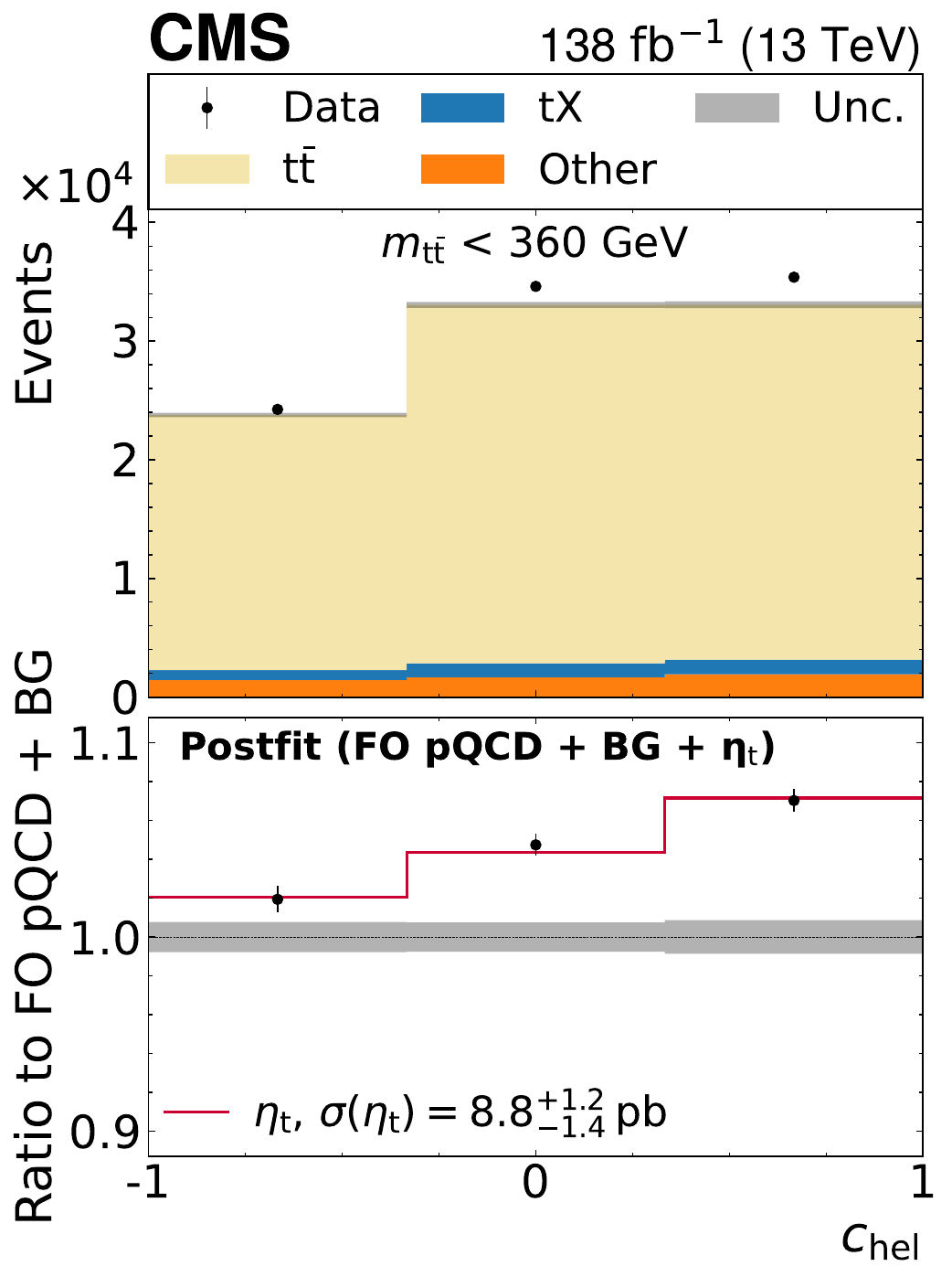}%
\hspace{0.1\linewidth}%
\includegraphics[width=0.32\textwidth]{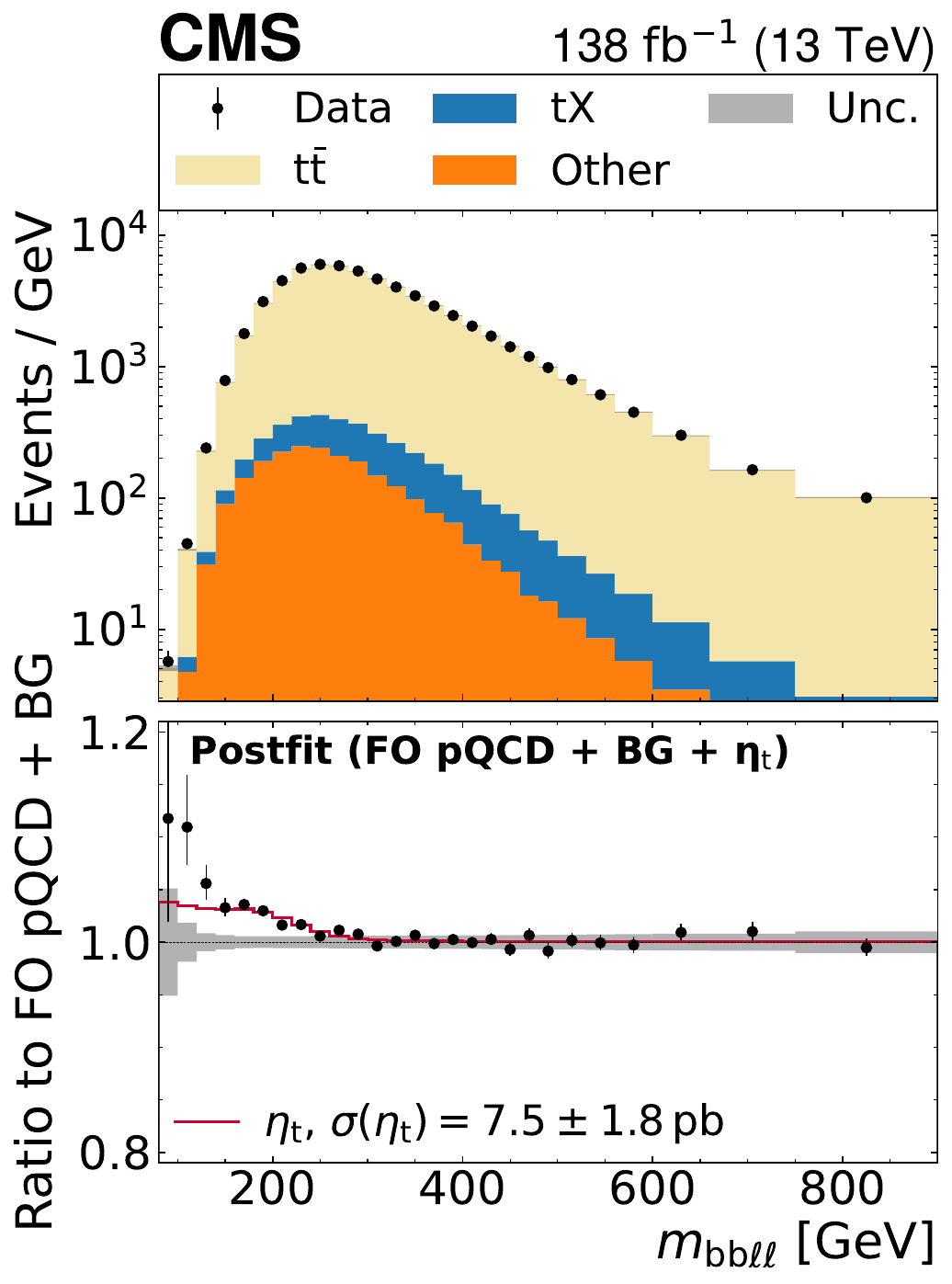}%
\caption{%
    Observed (points with statistical error bars) and predicted (stacked colored histograms) distributions.
    Left: \chel for $\mtt<360\GeV$ and integrated over \chan, from the nominal fit using \mtt.
    Right: \mbbll integrated over \chel and \chan, from the alternative fit using \mbbll instead of \mtt, which is discussed in Section~\ref{sec:exp_method_check}.
    In the upper panels, the \ttbar histogram shows the FO pQCD prediction after the fit to the data that includes the \etat signal model (whose contribution is not drawn).
    On the right, the shown event rates are divided by the bin width.
    The lower panels display the ratio of the data to the \FOpQCDbackground prediction, with \etat signal overlaid at its best fit \etat cross section (red line).
    The gray band indicates the postfit uncertainty.
    The first and last \mbbll bins include all events with reconstructed \mbbll below 100 and above 750\GeV, respectively, and the drawn bin width is used for the normalization in these bins.
}
\label{fig:chel_mbbll_ll}
\end{figure*}

We observe an excess of events in the data with respect to the FO pQCD prediction.
This excess occurs mainly at low \mtt, and its strength depends on \chel and \chan.
We analyze the \mtt distribution in 3$\times$3 equal-sized \chel and \chan bins, and show three of these bins in Fig.~\ref{fig:combined_ll}: the bin ($-1<\chel<-1/3$, $-1<\chan<-1/3$) where the \etat contribution is expected to be small (left), the intermediate bin ($-1/3<\chel<1/3$, $-1/3<\chan<1/3$) (middle), and the bin ($1/3<\chel<1$, $1/3<\chan<1$) where the \etat contribution is expected to be enhanced~(right).
In Fig.~\ref{fig:chel_mbbll_ll} (left), the \chel distribution is shown for $\mtt<360\GeV$ and integrated over \chan.
The data show a steeper slope than the FO pQCD prediction, as expected from an additional pseudoscalar contribution.
As can be seen in both figures, the \etat model can accommodate the excess.

We measure an \etat production cross section of
\begin{equation*}\label{eq:etatfitxsec}
\etatxsec = 8.8 \pm 0.5 \stat \,^{+1.1}_{-1.3} \syst\unit{pb} = \etatxsecvalunit.
\end{equation*}
Here, the statistical component of the uncertainty is estimated by fixing all nuisance parameters to their postfit values, while the systematic component is obtained as the quadratic difference between the total and the statistical uncertainty.
Our result is comparable to the magnitude of the theory estimate of 6.4\unit{pb}, which is obtained by fitting the results of an NRQCD calculation from Ref.~\cite{Sumino:2010bv} and subtracting the NLO+NLL pQCD prediction~\cite{Fuks:2021xje}, within the range of $338<\mtt<350\GeV$.
This is compatible with the recent NRQCD calculation in Ref.~\cite{Garzelli:2024uhe}.
At the time of writing this paper, neither of these theoretical predictions has uncertainty estimates attached.
The cross section measured in this work is without any explicit restrictions on the \mtt range.
The significance of the excess over the hypothesis with no \etat production exceeds five standard deviations (SDs).

\begin{figure}[!htp]
\centering
\ifthenelse{\boolean{cms@external}}{%
    \includegraphics[width=\linewidth]{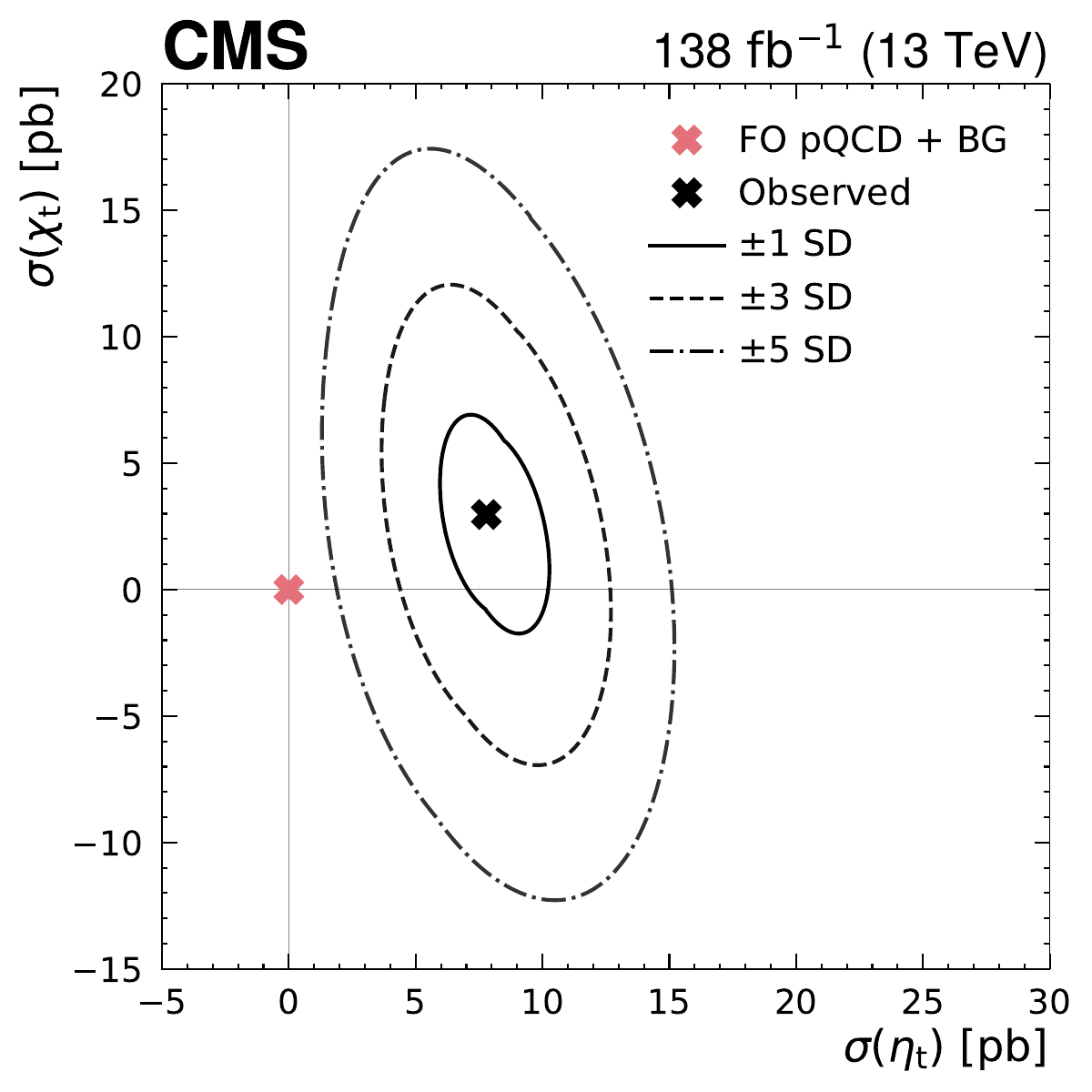}%
}{%
    \includegraphics[width=0.55\textwidth]{Figure_003.pdf}%
}
\caption{%
    Best fit value (cross) and allowed regions at one (solid line), three (dashed line), and five (dotted-dashed line) SDs for the cross section of \etat and \chit production, as observed in data (black). The \FOpQCDbackground expectation of zero \etat and \chit contributions is denoted by a pink star.
    Negative cross section values refer to a reduction of the \ttbar production cross sections with respect to the \FOpQCDbackground prediction around the threshold.
}
\label{fig:res_a_vs_h_contour}
\end{figure}

To explore the spin and \CP structure of the resonant excess, we perform an alternative fit that simultaneously includes contributions from pseudoscalar \etat and scalar \chit production.
The results are shown in Fig.~\ref{fig:res_a_vs_h_contour}.
While the data is compatible with zero contribution from the \chit state within one SD, zero contribution from the \etat state is excluded with a significance exceeding five SDs.
This establishes the presence of a pseudoscalar excess.

\section{Discussion}

\subsection{Robustness of the experimental analysis methods}
\label{sec:exp_method_check}

To ensure that the on-shell assumption used in the kinematic reconstruction algorithm does not sculpt the \mtt distribution to be signal-like through an unaccounted systematic effect, we repeat the analysis with a similar strategy but changing the fit distribution from \mtt to \mbbll.
This observable is defined without any assumption on the value of \mtop, and is shown in Fig.~\ref{fig:chel_mbbll_ll} (right).
To minimize dependence on modeling assumptions, nuisance parameters encoding the difference between the generators were not included in this check.
A cross section of $\etatxsec = 7.5 \pm 1.8 \unit{pb}$ is obtained in the \mbbll fit.
This is compatible with the nominal strategy, but with less sensitivity, which is expected.
We conclude that the kinematic reconstruction algorithm does not introduce spurious effects in our analysis.

As an additional check of the experimental reconstruction, we investigate the impact of possible jet \pt scale mismodeling on \mtt by fitting pseudodata obtained from the FO pQCD prediction with the jet \pt scale shifted up or down by its prefit uncertainty.
We find the resulting shift in the \etat cross section to be less than one SD of the nominal result.

\subsection{Comparison between fits with and without \texorpdfstring{\etat}{toponium} contribution}

\begin{figure*}[!htp]
\centering
\includegraphics[width=0.32\textwidth]{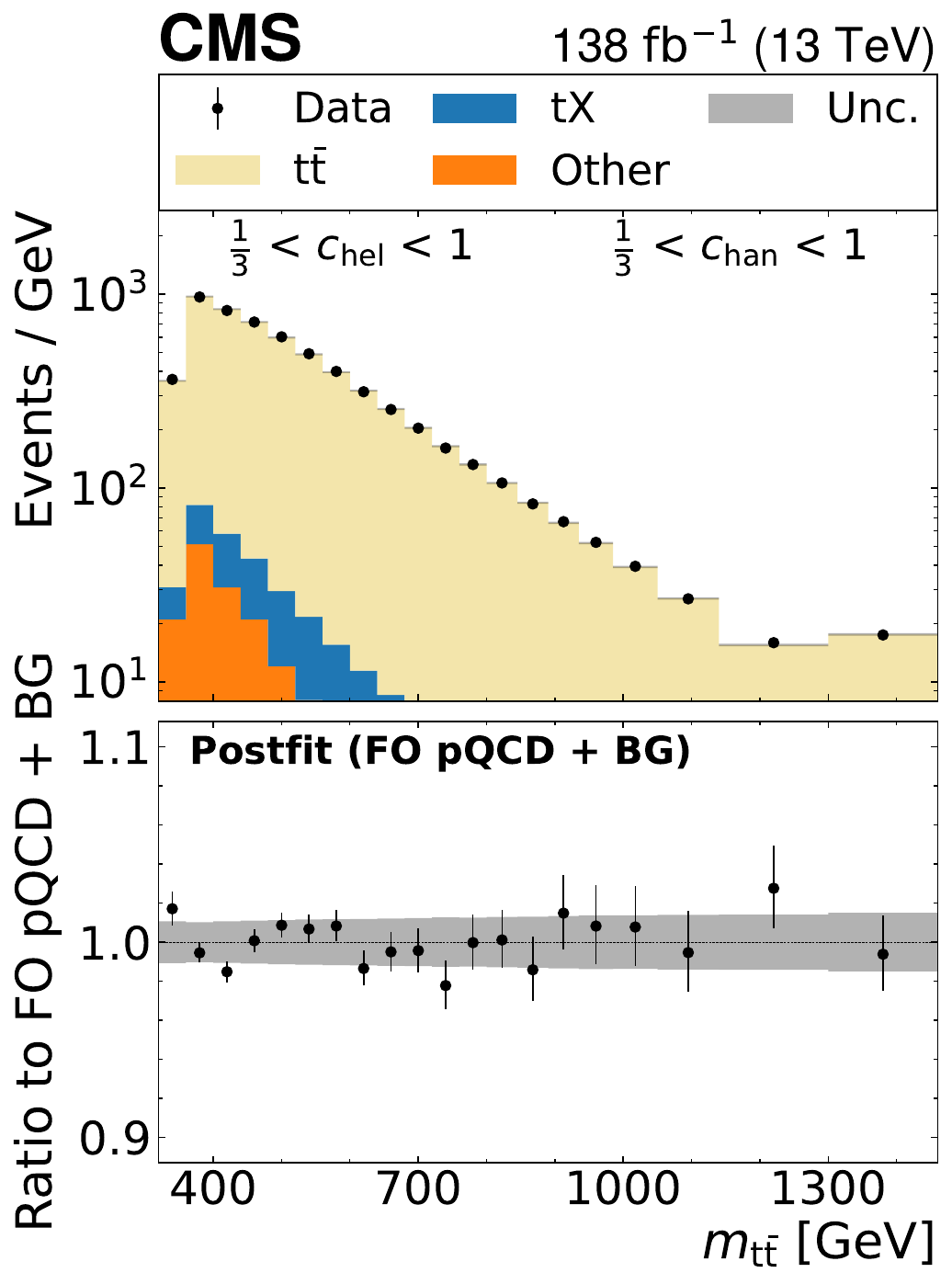}%
\hspace*{0.1\linewidth}%
\includegraphics[width=0.32\textwidth]{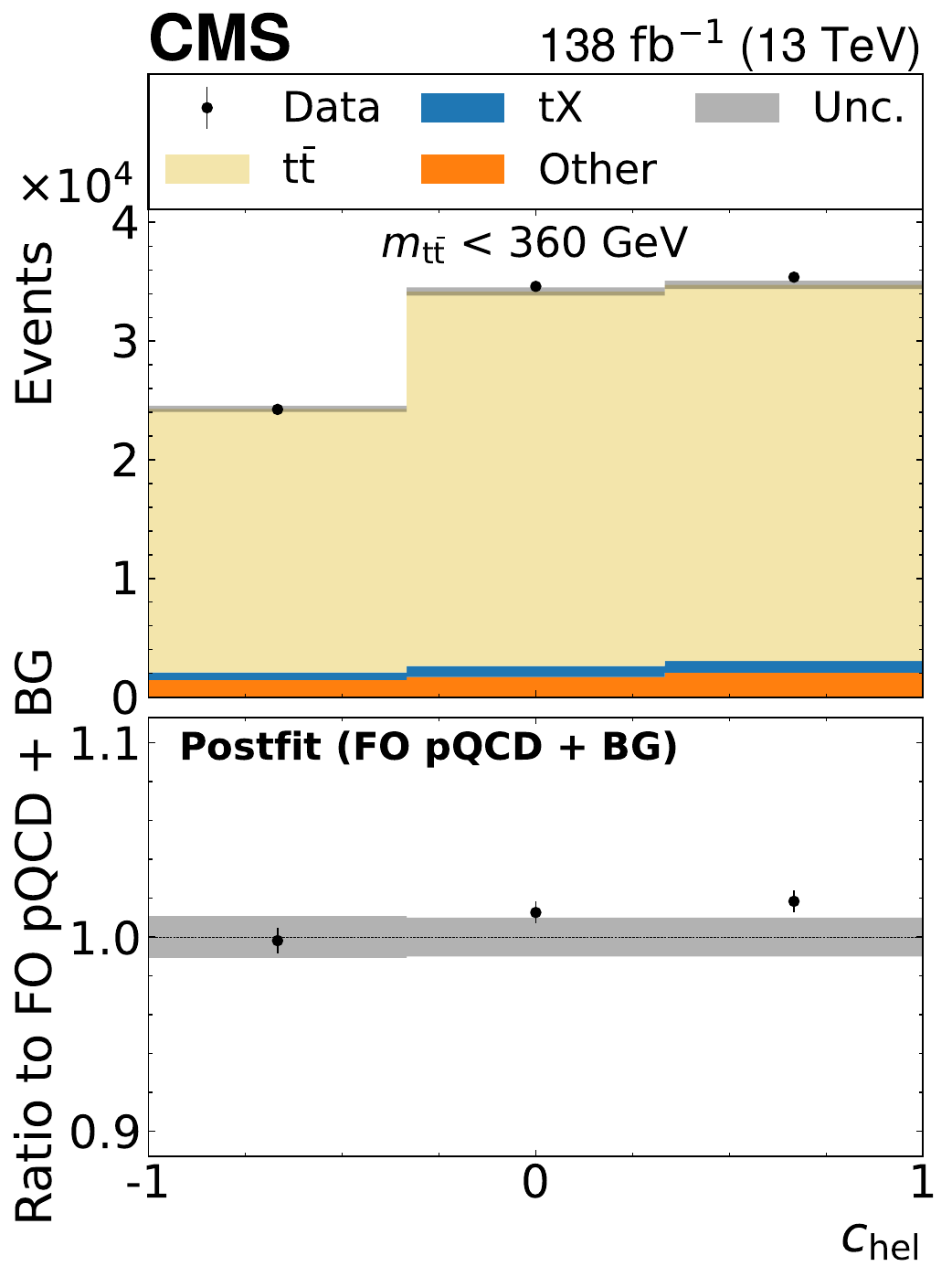}%
\caption{%
    Observed (points with statistical error bars) and predicted (stacked colored histograms) \mtt distribution in the (\chel, \chan) bin with the highest expected \etat contribution (left) and \chel distribution for $\mtt<360\GeV$ and integrated over \chan (right).
    In the upper panels, the histograms account only for the \FOpQCDbackground prediction, and are shown after the background-only fit to the data.
    On the left, the shown event rates are divided by the bin width.
    The lower panels display the ratio of the data to the \FOpQCDbackground prediction.
    The gray band indicates the postfit uncertainty.
    The binning is the same as in Figs.~\ref{fig:combined_ll} and \ref{fig:chel_mbbll_ll} (left).
}
\label{fig:bgonlypost}
\end{figure*}

The \mtt and \chel distributions obtained with a ``background-only'' fit, \ie, considering only the FO pQCD prediction plus background contributions without the \etat signal (labeled \FOpQCDbackground) and including all nuisance parameters, are shown in Fig.~\ref{fig:bgonlypost}.
The fit is able to absorb parts of the low-\mtt excess of data over prediction.
Nevertheless, a \chel distribution with a higher slope than the FO pQCD expectation remains visible, as expected for an \etat contribution, cf.\ Fig.~\ref{fig:chel_mbbll_ll} (left).

\begin{figure*}[!tp]
\centering
\includegraphics[width=0.85\textwidth]{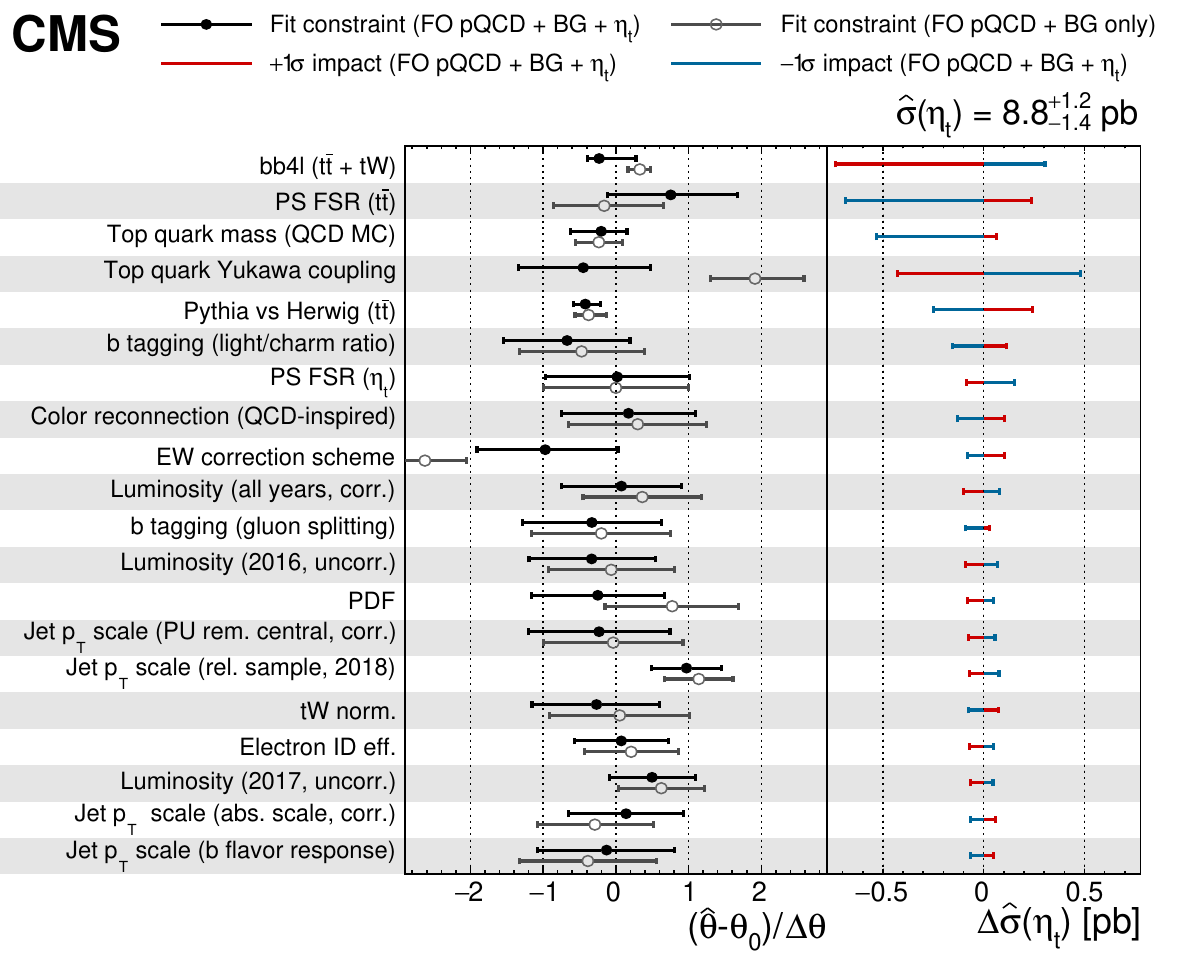}%
\caption{%
    For the nuisance parameters listed in the left column, the pulls $(\hat{\theta}-\theta_0)/\Delta\theta$ (middle column), where $\hat{\theta}$ and $\theta_0$ are the postfit and prefit values of the nuisance parameters and $\Delta\theta$ is the prefit uncertainty, are shown for the \FOpQCDbackgroundetat (black filled circles) and \FOpQCDbackground (gray empty circles) fit, as well as the impacts $\Delta\hat{\sigma}(\etat)$ (right column) for the \FOpQCDbackgroundetat fit.
    The impact $\Delta\hat{\sigma}(\etat)$ for a nuisance parameter $\theta$ is calculated by varying $\theta$ by $\pm$1 SD and evaluating the shift in \etatxsec.
    The nuisance parameters are ordered by the maximum of their $\pm$1 SD impacts in the \FOpQCDbackgroundetat fit.
}
\label{fig:impacts}
\end{figure*}

The quality of both the \FOpQCDbackgroundetat and \FOpQCDbackground fits is compared in Fig.~\ref{fig:impacts} through the postfit deviations of nuisance parameters from their prefit values (``pulls'') as well as their postfit reduction in uncertainty (``constraints'') in both cases.
In the \FOpQCDbackgroundetat fit (black filled circles), no nuisance parameter is pulled from its prefit value by more than one SD.
In contrast, in the \FOpQCDbackground fit (gray empty circles), the two nuisance parameters associated with the EW corrections (the top quark Yukawa coupling and the correction scheme) are pulled strongly away from their expected values. Since the EW contribution from diagrams containing a Higgs boson is expected to cause enhancements at the \ttbar threshold similar to \etat~\cite{Kuhn:2013zoa}, this can compensate for the observed excess of events and slope in \chel.
This, together with the fact that there is still a residual positive slope near the \ttbar threshold in Fig.~\ref{fig:bgonlypost}, corroborates the conclusion of the statistical analysis that a nonzero \etat contribution is the more physically sensible interpretation of the data.

Other significant nuisance parameters are constrained in both the \FOpQCDbackgroundetat and the \FOpQCDbackground fit.
These include the top quark mass, to which \mtt is naturally sensitive, and the differences between \PYTHIA and \HERWIG as well as between \hvq and \bbfourl, which are discussed in the next section.
We investigate whether the observed pulls and constraints could be attributed to imperfect modeling in specific regions of phase space by splitting each relevant nuisance parameter in two different ways: by binning separately in \chel and \chan, or by dividing into three \mtt regions $\mtt<400\GeV$, $400<\mtt<600\GeV$, and $\mtt>600\GeV$.
In all cases, the resulting constraints are comparable with those obtained in the nominal fit.
Furthermore, we verify that the prefit uncertainty ranges of these nuisance parameters do not significantly impact the final result by performing alternative fits in which the modeling-related nuisance parameters are left unconstrained.
The resulting \etat cross sections remain compatible with those from the nominal fit, even when the nuisance parameters are split as described above.

\subsection{Comparison between different baseline FO pQCD predictions}
\label{sec:gen_compare}

Here, we present and discuss the comparison between different setups to generate the \FOpQCDbackground predictions.
Fit results obtained with the nominal FO pQCD \ttbar prediction and the three alternative generator setups introduced in Section~\ref{sec:samples} are compared in Table~\ref{tab:generators}.
The setup ``\POWHEGhvqPYTHIA'' refers to the sample used in the nominal result.
The fit results shown here do not include the two systematic uncertainties for the choice of the ME event generator and the parton shower program, which are evaluated from comparisons between the different generator setups.
The setup ``\POWHEGbbfourlPYTHIA'' does not use the nominal \tW prediction and has dedicated samples to derive systematic uncertainty templates, whereas the other three setups use the same nominal \tW prediction and share the samples used to derive systematic uncertainty templates with the nominal result.
The largest difference in \etatxsec with respect to the nominal result is found for \POWHEGbbfourlPYTHIA, which is lower by $\sim$1.5 SDs.

\begin{table}[!ht]
\centering
\topcaption{%
    Results for \etatxsec obtained with different simulated event samples for the \ttbar FO pQCD (+\tW) prediction.
    Nuisance parameters encoding the difference between different generators are not included in these results.
    The nominal result, \ie, \POWHEGhvqPYTHIA including these nuisance parameters, is shown for comparison.
}
\renewcommand{\arraystretch}{1.1}
\cmsTable{\begin{tabular}{cc}
    \hline
    FO pQCD generator setup & \etatxsec [pb]{\Large\strut} \\
    \hline
    \POWHEGhvqPYTHIA & $8.7 \pm 1.1$ \\
    \POWHEGhvqHERWIG & $8.6 \pm 1.1$ \\
    \MGvATNLOFXFXPYTHIA & $9.8 \pm 1.3$ \\
    \POWHEGbbfourlPYTHIA & $6.6 \pm 1.4$ \\[\cmsTabSkip]
    Nominal result & \etatxsecval{\Large\strut} \\
    \hline
\end{tabular}}
\label{tab:generators}
\end{table}

\begin{figure*}[!htp]
\centering
\includegraphics[width=0.32\textwidth]{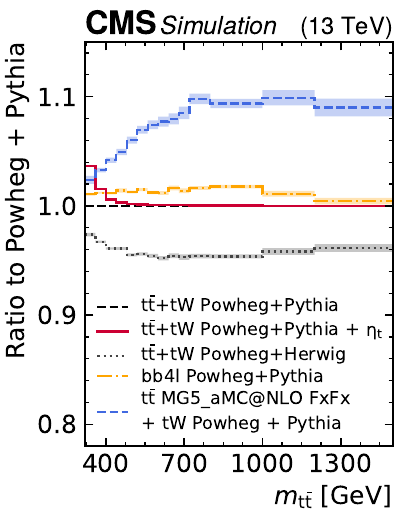}%
\hfill%
\includegraphics[width=0.32\textwidth]{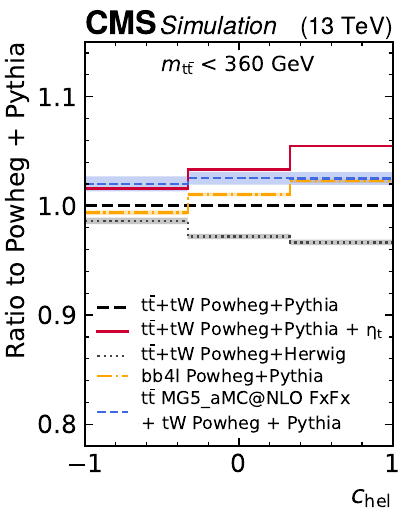}%
\hfill%
\includegraphics[width=0.32\textwidth]{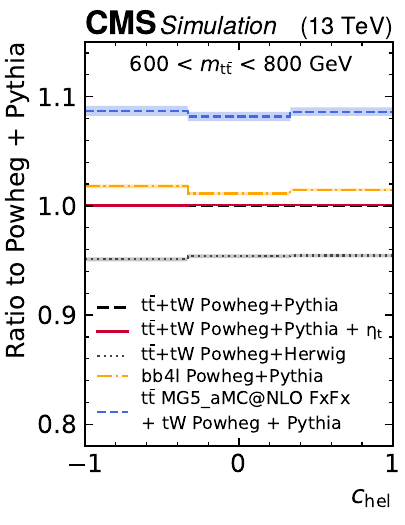}%
\caption{%
    Ratios of the predictions for \POWHEGhvqHERWIG (gray), \POWHEGbbfourlPYTHIA (orange), and \MGvATNLOFXFXPYTHIA (blue) to \POWHEGhvqPYTHIA (black) for \mtt (left), \chel at the \ttbar threshold ($\mtt<360\GeV$, center), and \chel in the \ttbar continuum ($600<\mtt<800\GeV$, right), both integrated over \chan.
    The effect of adding \etat to \POWHEGhvqPYTHIA is shown in red for comparison.
    The shaded bands denote the statistical uncertainties from the MC simulation.
}
\label{fig:generators}
\end{figure*}

To illustrate the differences between the generator setups, the ratios to the nominal \POWHEGhvqPYTHIA prediction are shown in Fig.~\ref{fig:generators} for the \mtt distribution, as well as the \chel distribution at the \ttbar threshold and in the \ttbar continuum.
It can be seen that \HERWIG, compared to \PYTHIA, predicts a lower \ttbar acceptance, an increase of events at low \mtt (similar to \etat), and a reduced slope in \chel for low \mtt (opposite to \etat). Since these effects are contradicting the prediction from \etat production, the observed \etatxsec value remains basically unchanged.
In contrast, the \bbfourl sample predicts a stronger slope in \chel for low \mtt, which is similar to \etat and thus explains the smaller observed \etatxsec value when this setup is used as the background prediction.
Fig.~\ref{fig:generators} (right) shows, however, that above the threshold the considered generators predict the same slope in \chel.

These findings, as presented in Table~\ref{tab:generators},
are consistent with the constraints and impacts of the two nuisance parameters associated with the differences between \PYTHIA and \HERWIG as well as between \hvq and \bbfourl, shown in Fig.~\ref{fig:impacts}.
In particular, the strong impact of \bbfourl on the extracted value of \etatxsec is apparent.
Notably, in Fig.~\ref{fig:impacts}, the nuisance parameter associated with \bbfourl is constrained around zero, indicating that the nominal \hvq prediction provides a better description of the data.
A detailed explanation of this observation is left for future investigations.

Furthermore, \MGvATNLO predicts a significant slope in \mtt compared to \POWHEG, while the spin correlations are unchanged.
This is consistent with an increase in the extracted value of \etatxsec, as given in Table~\ref{tab:generators}.

\section{Summary}

We report the observation of resonant top quark-antiquark (\ttbar) production near the kinematic production threshold, with spin properties consistent with contributions from a pseudoscalar state,
using proton-proton collision data recorded by the CMS experiment at $\sqrt{s}=13\TeV$ in 2016--2018 and corresponding to an integrated luminosity of 138\fbinv.
The data is compared to the standard model prediction including only nonresonant \ttbar production obtained with fixed-order calculations in perturbative quantum chromodynamics.
An excess is observed with respect to this model, with a statistical significance exceeding five standard deviations.
We emphasize, however, that the modeling of the \ttbar threshold region is challenging and requires further theoretical investigation.
It is worth noting that alternative explanations of the excess are also plausible, given the current experimental resolution of the \ttbar invariant mass; we explore this direction further in Ref.~\cite{CMS:HIG-22-013}.
The result is compatible with the formation of a \oneSone toponium quasi-bound state \etat with a measured cross section of $\etatxsec=\etatxsecvalunit$, based on a simplified model inspired by nonrelativistic quantum chromodynamics.

\begin{acknowledgments}
We congratulate our colleagues in the CERN accelerator departments for the excellent performance of the LHC and thank the technical and administrative staffs at CERN and at other CMS institutes for their contributions to the success of the CMS effort. In addition, we gratefully acknowledge the computing centers and personnel of the Worldwide LHC Computing Grid and other centers for delivering so effectively the computing infrastructure essential to our analyses. Finally, we acknowledge the enduring support for the construction and operation of the LHC, the CMS detector, and the supporting computing infrastructure provided by the following funding agencies: SC (Armenia), BMBWF and FWF (Austria); FNRS and FWO (Belgium); CNPq, CAPES, FAPERJ, FAPERGS, and FAPESP (Brazil); MES and BNSF (Bulgaria); CERN; CAS, MoST, and NSFC (China); MINCIENCIAS (Colombia); MSES and CSF (Croatia); RIF (Cyprus); SENESCYT (Ecuador); ERC PRG, RVTT3 and MoER TK202 (Estonia); Academy of Finland, MEC, and HIP (Finland); CEA and CNRS/IN2P3 (France); SRNSF (Georgia); BMBF, DFG, and HGF (Germany); GSRI (Greece); NKFIH (Hungary); DAE and DST (India); IPM (Iran); SFI (Ireland); INFN (Italy); MSIP and NRF (Republic of Korea); MES (Latvia); LMTLT (Lithuania); MOE and UM (Malaysia); BUAP, CINVESTAV, CONACYT, LNS, SEP, and UASLP-FAI (Mexico); MOS (Montenegro); MBIE (New Zealand); PAEC (Pakistan); MES and NSC (Poland); FCT (Portugal);  MESTD (Serbia); MICIU/AEI and PCTI (Spain); MOSTR (Sri Lanka); Swiss Funding Agencies (Switzerland); MST (Taipei); MHESI and NSTDA (Thailand); TUBITAK and TENMAK (Turkey); NASU (Ukraine); STFC (United Kingdom); DOE and NSF (USA).

\hyphenation{Rachada-pisek} Individuals have received support from the Marie-Curie program and the European Research Council and Horizon 2020 Grant, contract Nos.\ 675440, 724704, 752730, 758316, 765710, 824093, 101115353, 101002207, and COST Action CA16108 (European Union); the Leventis Foundation; the Alfred P.\ Sloan Foundation; the Alexander von Humboldt Foundation; the Science Committee, project no. 22rl-037 (Armenia); the Fonds pour la Formation \`a la Recherche dans l'Industrie et dans l'Agriculture (FRIA-Belgium); the Beijing Municipal Science \& Technology Commission, No. Z191100007219010 and Fundamental Research Funds for the Central Universities (China); the Ministry of Education, Youth and Sports (MEYS) of the Czech Republic; the Shota Rustaveli National Science Foundation, grant FR-22-985 (Georgia); the Deutsche Forschungsgemeinschaft (DFG), among others, under Germany's Excellence Strategy -- EXC 2121 ``Quantum Universe" -- 390833306, and under project number 400140256 - GRK2497; the Hellenic Foundation for Research and Innovation (HFRI), Project Number 2288 (Greece); the Hungarian Academy of Sciences, the New National Excellence Program - \'UNKP, the NKFIH research grants K 131991, K 133046, K 138136, K 143460, K 143477, K 146913, K 146914, K 147048, 2020-2.2.1-ED-2021-00181, TKP2021-NKTA-64, and 2021-4.1.2-NEMZ\_KI-2024-00036 (Hungary); the Council of Science and Industrial Research, India; ICSC -- National Research Center for High Performance Computing, Big Data and Quantum Computing and FAIR -- Future Artificial Intelligence Research, funded by the NextGenerationEU program (Italy); the Latvian Council of Science; the Ministry of Education and Science, project no. 2022/WK/14, and the National Science Center, contracts Opus 2021/41/B/ST2/01369 and 2021/43/B/ST2/01552 (Poland); the Funda\c{c}\~ao para a Ci\^encia e a Tecnologia, grant CEECIND/01334/2018 (Portugal); the National Priorities Research Program by Qatar National Research Fund;  MICIU/AEI/10.13039/501100011033, ERDF/EU, "European Union NextGenerationEU/PRTR", and Programa Severo Ochoa del Principado de Asturias (Spain); the Chulalongkorn Academic into Its 2nd Century Project Advancement Project, and the National Science, Research and Innovation Fund via the Program Management Unit for Human Resources \& Institutional Development, Research and Innovation, grant B39G670016 (Thailand); the Kavli Foundation; the Nvidia Corporation; the SuperMicro Corporation; the Welch Foundation, contract C-1845; and the Weston Havens Foundation (USA).
\end{acknowledgments}

\bibliography{auto_generated}
\cleardoublepage \appendix\section{The CMS Collaboration \label{app:collab}}\begin{sloppypar}\hyphenpenalty=5000\widowpenalty=500\clubpenalty=5000\input{TOP-24-007-public-authorlist.tex}\end{sloppypar}
%%% END EDITABLE REGION %%%
% skeleton_end
\end{document}

%% file: TOP-24-007-public-authorlist.tex
\cmsinstitute{Yerevan Physics Institute, Yerevan, Armenia}
{\tolerance=6000
A.~Hayrapetyan, V.~Makarenko\cmsorcid{0000-0002-8406-8605}, A.~Tumasyan\cmsAuthorMark{1}\cmsorcid{0009-0000-0684-6742}
\par}
\cmsinstitute{Institut f\"{u}r Hochenergiephysik, Vienna, Austria}
{\tolerance=6000
W.~Adam\cmsorcid{0000-0001-9099-4341}, J.W.~Andrejkovic, L.~Benato\cmsorcid{0000-0001-5135-7489}, T.~Bergauer\cmsorcid{0000-0002-5786-0293}, K.~Damanakis\cmsorcid{0000-0001-5389-2872}, M.~Dragicevic\cmsorcid{0000-0003-1967-6783}, C.~Giordano, P.S.~Hussain\cmsorcid{0000-0002-4825-5278}, M.~Jeitler\cmsAuthorMark{2}\cmsorcid{0000-0002-5141-9560}, N.~Krammer\cmsorcid{0000-0002-0548-0985}, A.~Li\cmsorcid{0000-0002-4547-116X}, D.~Liko\cmsorcid{0000-0002-3380-473X}, I.~Mikulec\cmsorcid{0000-0003-0385-2746}, J.~Schieck\cmsAuthorMark{2}\cmsorcid{0000-0002-1058-8093}, R.~Sch\"{o}fbeck\cmsAuthorMark{2}\cmsorcid{0000-0002-2332-8784}, D.~Schwarz\cmsorcid{0000-0002-3821-7331}, M.~Shooshtari, M.~Sonawane\cmsorcid{0000-0003-0510-7010}, W.~Waltenberger\cmsorcid{0000-0002-6215-7228}, C.-E.~Wulz\cmsAuthorMark{2}\cmsorcid{0000-0001-9226-5812}
% ...
\par}
\vskip\cmsinstskip
\dag:~Deceased\\
$^{1}$Also at Yerevan State University, Yerevan, Armenia\\
$^{2}$Also at TU Wien, Vienna, Austria\\
% ...
